\documentclass{emulateapj}

\usepackage{apjfonts}
\def\Mr{M_{^{0.1}r} < -21}
\def\wrp{w_p(r_p)}
\def\Mmin{M_{\rm min}}
\def\sigcen{\sigma_{\rm cen}}

\def\Ncen{\langle N_{\rm cen}(M)\rangle}
\def\Nsatavg{\langle N_{\rm sat}(M)\rangle}

\font\FermiSmallfont=cmssq8 scaled 1200

\def\LANLppthead#1#2#3{
\null 
\begin{center}\vskip -1.0truein{\hbox to 7.5truein {
\hfill
\vbox to 1in {\vfill \FermiSmallfont
              \hbox{#1}
              \hbox{#2}
              \hbox{#3}
              \vfill}
}}\vskip-0.0truein\end{center}}

\shorttitle{Cosmology and the HOD from Small-Scale Clustering}
\shortauthors{Abazajian et al.}

\begin{document}

\LANLppthead {LA-UR 04-1121}{FNAL-PUB-04-137-A}{astro-ph/0408003}

\title{Cosmology and the Halo Occupation Distribution from Small-Scale\\ Galaxy
  Clustering in the Sloan Digital Sky Survey}

\author{Kevork Abazajian\altaffilmark{1,2}, Zheng Zheng\altaffilmark{3}, Idit
   Zehavi\altaffilmark{4}, David H.\ Weinberg\altaffilmark{3}, Joshua A.\
   Frieman\altaffilmark{2,5}, \\Andreas A.\ Berlind\altaffilmark{6}, Michael R.\
   Blanton\altaffilmark{6},  Neta A.\ Bahcall\altaffilmark{7}, J.\
   Brinkmann\altaffilmark{8}, Donald P.\ Schneider\altaffilmark{9}, \\Max
   Tegmark\altaffilmark{10,11}}

\begin{abstract}
We use the projected correlation function $w_p(r_p)$ of a volume-limited
subsample of the Sloan Digital Sky Survey (SDSS) main galaxy redshift catalogue
to measure the halo occupation distribution (HOD) of the galaxies of the
sample. Simultaneously, we allow the cosmology to vary within cosmological
constraints imposed by cosmic microwave background experiments in a
$\Lambda$CDM model.  We find that combining $w_p(r_p)$ for this sample alone
with the observations by WMAP, ACBAR, CBI and VSA can provide one of the most
precise techniques available to measure cosmological parameters. For a minimal
flat six-parameter $\Lambda$CDM model with an HOD with three free parameters,
we find $\Omega_m=0.278^{+0.027}_{-0.027}$, $\sigma_8=0.812^{+0.028}_{-0.027}$,
and $H_0=69.8^{+2.6}_{-2.6}\rm\, km\, s^{-1}\, Mpc^{-1}$; these errors are
significantly smaller than from CMB alone and similar to those obtained by
combining CMB with the large-scale galaxy power spectrum assuming
scale-independent bias.  The corresponding HOD parameters describing the
minimum halo mass and the normalization and cut-off of the satellite mean
occupation are $\Mmin=(3.03^{+0.36}_{-0.36})\times 10^{12} h^{-1}\, M_\odot$,
$M_1 = (4.58^{+0.60}_{-0.60})\times 10^{13} h^{-1}\, M_\odot$, and
$\kappa=4.44^{+0.51}_{-0.69}$.  These HOD parameters thus have small fractional
uncertainty when cosmological parameters are allowed to vary within the range
permitted by the data.  When more parameters are added to the HOD model, the
error bars on the HOD parameters increase because of degeneracies, but the
error bars on the cosmological parameters do not increase greatly.  Similar
modeling for other galaxy samples could reduce the statistical errors on these
results, while more thorough investigations of the cosmology dependence of
nonlinear halo bias and halo mass functions are needed to eliminate remaining
systematic uncertainties, which may be comparable to statistical uncertainties.
\end{abstract}


\altaffiltext{1}{Theoretical Division, MS B285, Los Alamos
  National Laboratory, Los Alamos, NM 87545}
\altaffiltext{2}{Theoretical Astrophysics Group, 
  Fermi National Accelerator Laboratory, P.O. Box 500, Batavia,
  IL 60510}
\altaffiltext{3}{Department of Astronomy, Ohio State University, Columbus,
  OH 43210}
\altaffiltext{4}{Steward Observatory, University of Arizona, 
933 N.~Cherry Ave., Tucson, AZ 85721}
\altaffiltext{5}{Kavli Institute for Cosmological Physics, 
  Department of Astronomy and Astrophysics, The University of
  Chicago, 5640 S.~Ellis Ave., Chicago, IL 60637}
\altaffiltext{6}{Center for Cosmology and Particle Physics, Department of
  Physics, New York University, 4 Washington Place, New York, NY 10003}
\altaffiltext{7}{Apache Point Observatory, P.O. Box 59, Sunspot, NM 88349}
\altaffiltext{8}{Department of Astrophysical Sciences, Princeton University, 
Princeton, NJ 08544}
\altaffiltext{9}{Department of Astronomy and Astrophysics, 
The Pennsylvania State University, University Park, PA 16802}
\altaffiltext{10}{Department of Physics, University of Pennsylvania, Philadelphia, PA 19104}
\altaffiltext{11}{Department of Physics, Massachusetts Institute of Technology,
  Cambridge, MA 02139}
\section{Introduction}

Over the last several years, halo occupation models of galaxy bias have led to
substantial progress in characterizing the relation between the distributions
of galaxies and dark matter.  Gravitational clustering of the dark matter
determines the population of virialized dark matter halos, with essentially no
dependence on the more complex physics of the sub-dominant baryon component.
Galaxy formation physics determines the halo occupation distribution (HOD),
which specifies the probability $P(N|M)$ that a halo of virial mass $M$
contains $N$ galaxies of a given type, together with any spatial and velocity
biases of galaxies within halos
\citep{kauffmann97,benson00,berlind03,kravtsov04}.  Given cosmological
parameters and a specified HOD, one can calculate any galaxy clustering
statistic, on any scale, either by populating the halos of N-body simulations
(e.g., \citealt{jing98,jing02,berlind02}) or by using an increasingly powerful
array of analytic approximations (e.g.,
\citealt{peacock00,seljak00,scoccimarro01,takada03}; see \citealt{cooray02} for
a recent review).  

The 2dF Galaxy Redshift Survey (2dFGRS) \citep{Colless:2003wz} and the Sloan
Digital Sky Survey (SDSS) \citep{York:2000gk,abazajian03,abazajian04} allow
galaxy clustering measurements of unprecedented precision and detail, making
them ideal data sets for this kind of modeling.  \citet{zehavi04} (hereafter
Z04a) show that the projected correlation function $w_p(r_p)$ of luminous
($\Mr$) SDSS galaxies exhibits a statistically significant departure from a
power law, and that a 2-parameter HOD model applied to the prevailing
$\Lambda$CDM (cold dark matter with a cosmological constant) cosmology accounts
naturally for this departure, reproducing the observed $w_p(r_p)$.  Here,
$M_{^{0.1}r}$ is the absolute magnitude in the redshifted $r$ band, with
observed magnitudes K-corrected to rest frame magnitudes for the SDSS bands
blueshifted by $z=0.1$, the median redshift of the survey
\citep{blanton03}. \cite{magliocchetti03} have applied a similar type of
analysis to $\wrp$ for a fixed cosmology in the 2dFGRS.  The halo model plus
HOD has also been used to successfully describe the clustering of Lyman-break
galaxies \citep{porciani2002}, high-redshift red galaxies \citep{zheng04}, as
well as 2dF quasars \citep{porciani04}. Recently, it was shown that large-scale
overdensities are not correlated with galaxy color or star formation history at
a fixed small-scale overdensity, supporting the HOD ansatz that a galaxy's
properties are related only to the host halo mass and not the large-scale
environment \citep{Blanton:2004an}.

In this paper, we go beyond the Z04a analysis by bringing in additional
cosmological constraints from cosmic microwave background (CMB) measurements
and allowing the HOD and cosmological parameters to vary simultaneously.  This
investigation complements that of \citet{zehavi04-2} (hereafter Z04b), who
examine the luminosity and color dependence of galaxy HOD parameters for a
fixed cosmology.  It also complements analyses that combine CMB data with the
large-scale {\it power spectrum} measurements from the 2dFGRS or SDSS (e.g.,
\citealt{percival01,spergel03,tegmark04}).  Such analyses use linear
perturbation theory to predict the dark matter power spectrum, and they assume
that galaxy bias is scale-independent in the linear regime.  It also
complements HOD and cosmological parameter determination approaches using
galaxy-galaxy lensing in the SDSS \citep{seljaklens04} and their combination
with Lyman-$\alpha$ forest clustering in the SDSS quasar sample
\citep{seljakparams04}.  Our analysis draws on data that extend into the highly
non-linear regime, and in place of scale-independent bias it adopts a
parameterized form of the HOD motivated by theoretical studies of galaxy
formation.

\section{Theory}
\label{theory}

We explore spatially flat, ``vanilla'' cosmological models that have six
parameters, $(\Omega_b h^2, \Omega_c h^2,\Theta_s,\ln(A),n,\tau)$, where
$\Omega_b$ and $\Omega_c$ are fractions of the critical density in baryons and
cold dark matter; $\Theta_s$ is the angular acoustic peak scale of the CMB, a
useful proxy for the Hubble parameter, $H_0 = 100\, h
\rm\ km\;s^{-1}\,Mpc^{-1}$; $A$ and $n$ are the amplitude and tilt of the
primordial scalar fluctuations; $\tau$ is the optical depth due to
reionization.  

In the halo model of galaxy clustering, the two-point correlation function of
galaxies is composed of two parts, the 1-halo term and the 2-halo term, $\xi(r)
= 1+\xi_{\rm 1h}(r)+\xi_{\rm 2h}(r)$, which represent contributions by galaxy
pairs from same halos and different halos which dominate at small scales and
large scales, respectively.  Here, the correlation function is calculated at
the effective redshift of our observed SDSS sample at $z=0.1$, which is a
nontrivial modification since errors on the amplitude of the power spectrum at
small scales (i.e., $\sigma_8$) are found be comparable to the growth factor
shift at $z=0.1$ (see below).  The calculation of the 1-halo term is
straightforward (e.g., \citealt{berlind02}):
\begin{eqnarray}
1+\xi_{\rm 1h}(r) =
\frac{1}{2\pi r^2\bar{n}^2_g} \int_0^\infty &&{\frac{dn}{dM} 
\frac{\langle N(N-1) \rangle_M}{2}}\nonumber\\
 && \times  \frac{1}{2 R_{\rm vir}(M)}F^\prime \left(\frac{r}{2
        R_{\rm vir}}\right)\,dM,
\end{eqnarray}
where $\bar n_g$ is the mean number density of galaxies calculated from the HOD
and halo model, $dn/dM$ is the halo mass function, $\langle N(N-1) \rangle_M/2$
is the average number of galaxy pairs in a halo of mass $M$, and $F(r/2 R_{\rm
  vir})$ is the cumulative radial distribution of galaxy pairs
\citep{berlind02,zheng04}.

For the 2-halo term term, in order to reach the accuracy needed to model the
SDSS data, we include the nonlinear evolution of matter clustering and the halo
exclusion effect.  This is done in Fourier space,
\begin{equation}
\xi_{\rm 2h}(r) = \frac{1}{2\pi^2}\int_0^\infty{P_{\rm gg}^{\rm
    2h}(k)k^2\frac{\sin kr}{kr}\, dk},
\end{equation} 
where
\begin{equation}
P_{\rm gg}^{\rm 2h}(k) = P_{mm}^{\rm NL}(k)\left[\frac{1}{\bar
    n_g}\int_0^{M_{\rm max}}{dM\, \frac{dn}{dM} \langle N(M)\rangle b_h(M)
    y_g(k,M)}\right]^2.
\label{pgg}
\end{equation} 
The mean occupation of halos of mass $M$ is $\langle N(M)\rangle$, $y_g(k,M)$
is the normalized Fourier transform of the galaxy distribution profile in a
halo of mass $M$.  We approximate halo exclusion effects in two-halo
correlation separations of $r$ by choosing the upper limit of the integral in
Eq.~(\ref{pgg}) such that $M_{\rm max}$ is the mass of a halo with virial
radius $r/2$, as incorporated in \citet{zheng04}, Z04a, and Z04b.  The
importance of the nonlinear matter power spectrum and halo exclusion in
accurately modeling the two-halo galaxy correlation function was also found by
\citet{magliocchetti03} and \citet{Wang:2004km}.

In order to accurately include the dependence of the halo modeling of galaxy
clustering for a varying cosmology, we include cosmologically general (within
$\Lambda$CDM) forms of the nonlinear matter power spectrum, halo bias, halo
mass function, and dark matter halo concentration.  We use the nonlinear matter
spectrum $P_{mm}^{\rm NL}(k)$ of Smith et al.'s (2003) {\tt halofit} code,
modified to utilize a numerically calculated transfer function from the Code
for Anisotropies in the Microwave Background (CAMB, \citealt{lewis00}), based
on CMBFAST \citep{seljak96}.  We use halo bias factors $b_h(M)$ determined in
the high-resolution simulations of \citet{seljak04}, along with its given
cosmological dependence, which provides a better fit (lower $\chi^2$) to our
observational data than halo bias models based on the peak background split
\citep{sheth99,sheth01}.  We use the \citet{jenkins01} spherical overdensity of
180 [$\rm SO(180)$, Eq.~(B3)] halo mass function, and include in its
interpretation of the definition of halo mass the variation in the virial
overdensity with cosmology \citep{Bryan:1998dn}
\begin{equation}
\Delta_v = \frac{18\pi^2+82x-39x^2}{1+x},
\end{equation}
and its effect in relating the \citet{jenkins01} mass function to varying
cosmologies (see, e.g., \citealt{Whitehalo01,Hu:2002we}). Here,
$x\equiv\Omega_m(z)-1$.  The variation of the virial overdensity also changes
the halo exclusion scale of $R_{\rm vir}(M_{\rm max})$ used in Eq.~(\ref{pgg}).
Since our luminous SDSS subsample populates halos of $M>10^{12}M_\odot$, the
breakdown of the \citet{jenkins01} mass-function fit at $M\lesssim
10^{10}M_\odot$ is not important.

We assume that the average spatial distribution of satellite galaxies within a
halo follows a Navarro-Frenk-White (NFW) density profile of the dark matter
\citep{Navarro:1996iw}, motivated by hydrodynamic simulation results
\citep{white01,berlind03} and N-body simulation galaxy clustering predictions
with halos populated by semi-analytic models
\citep{kauffmann97,kauffmann99,benson00,somerville01}.  However, as a test, we
drop this assumption of no spatial bias within halos between galaxies and dark
matter and find it is not important (see \S\ref{section:results} below).
In the case of no spatial bias, each halo is assumed to have a
cosmologically-dependent concentration
\begin{equation}
c=c_0\left(\frac{M}{M_*}\right)^\beta,
\label{conc}
\end{equation}
where
\begin{eqnarray}
c_0 &=& 11\left(\frac{\Omega_m}{0.3}\right)^{-0.35}\left(\frac{n_{\rm
    eff}}{-1.7}\right)^{-1.6},\\
\beta &=&-0.05,
\label{c0}
\end{eqnarray}
as found in fits to numerical results for varying
cosmologies by \citet{huffenberger03}.  Here,
\begin{equation}
n_{\rm eff} \equiv \frac{d\ln P_{\rm lin}(k)}{d\ln k}\Big|_{k_*},
\end{equation}
where $k_*$ is the nonlinear scale such that $\Delta^2_{\rm lin}(k_*)=1$, $M$
is the virial mass of the halo and $M_*$ is the nonlinear mass scale.  There is
a scatter about any mean concentration value and this could change the
prediction of the shape of a given halo.  However, as we describe below, our
results are largely insensitive to the exact form of the concentration of the
galaxies with respect to the dark matter.

Our HOD parameterization for a lu\-mi\-no\-si\-ty-thre\-shold galaxy sample
($\Mr$ in this paper) is motivated by results of substructures
from high-resolution dissipationless simulations of \citet{kravtsov04}.  The
HOD has a simple form when separated into central and satellite galaxies.  The
mean occupation number of central galaxies is modeled as a step function at
some minimum mass, smoothed by a complementary error function such that
\begin{equation}
\Ncen=\frac{1}{2}{\rm Erfc}\left[\frac{\ln(M_{\rm
min}/M)}{\sqrt{2}\sigcen}\right],
\end{equation} 
to account for scatter in the relation between the adopted magnitude limit and
the halo mass limit \citep{zheng04b}.  (Note that the number of central
galaxies is always $N_{\rm cen}=0$ or $N_{\rm cen}=1$ by definition.) The
occupation number of satellite galaxies is well approximated by a Poisson
distribution with the mean following a power law,
\begin{equation}
\Nsatavg=\left[\frac{M-\kappa \Mmin}{M_1}\right]^\alpha,
\end{equation}
where we introduce a smooth cut-off of the average satellite number at a
multiple $\kappa \ge 1$ of the minimum halo mass.  \citet{guzik02} and
\citet{berlind03}, using semi-analytic model calculations, and
\cite{kravtsov04}, using high resolution $N$-body simulations, found $\alpha
\approx 1$.  The general HOD above is characterized by five quantities:
$\Mmin$, $M_1$, $\sigcen$, $\kappa$, and $\alpha$, and we refer to this as the
$5p$ model.  It provides an excellent fit to predictions of semi-analytic
models and hydrodynamic simulations \citep{zheng04b}, in addition to describing
subhalo populations in N-body simulations \citep{kravtsov04}.  

\section{Observations}
\label{section:observations}
The SDSS uses a suite of specialized instruments and data reduction pipelines
\citep{Gunn:1998vh,Hogg:2001gc,Pier:2002iq,Stoughton:2002ae} to image the sky
in five passbands \citep{Fukugita:1996qt,Smith:2002pc} and obtain spectra of
well defined samples of galaxies and quasars
\citep{Eisenstein:2001cq,Richards:2002bb,Strauss:2002dj,blanton03-2}.  For our
analysis, we use Z04b's measurement of the projected correlation function
$\wrp$ of a volume-limited sample of galaxies with $\Mr$.  This sample is in
turn selected from a well characterized subset of the main galaxy sample as of
July, 2002, known as Large Scale Structure (LSS) {\tt sample12}, which includes
$\sim$200,000 galaxies over $2500\rm\, deg^2$ of sky \citep{Blanton:2004aa}. We
use the $\Mr$ sample with the full $\wrp$ data covariance matrix from
the jackknife estimates of Z04b.  There are 26,015 galaxies in the $\Mr$
sample.

\begin{figure}
\plotone{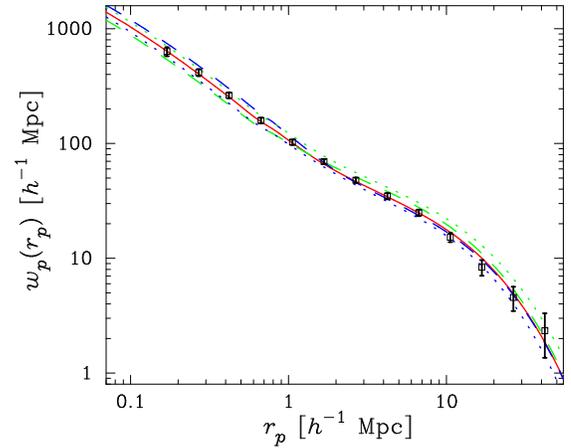}
\caption{Shown are the projected correlation function $w_p(r_p)$ of $\Mr$
  galaxies from the SDSS LSS {\tt sample12} (points with $1\sigma$ diagonal
  errors) and the best fit three parameter HOD model (solid).  Only points with
  $r<20h^{-1}\rm Mpc$ are used in the fit.  Also shown are predicted $w_p(r_p)$
  models with $\Omega_c$ (dotted) and $\sigma_8$ (dashed) at $\pm 3\sigma$ from
  their best fit values.  HOD parameters and other cosmological parameters are
  held fixed.  As seen here, the sensitivity to $\sigma_8$ and $\Omega_c$ comes
  from both the amplitude and combined shape of the 1-halo and 2-halo regimes
  of $w_p(r_p)$. \label{wpobs}}
\end{figure}

The observed projected correlation function is obtained from the 2-d
correlation function $\xi(r_p,\pi)$ by integrating along the line of sight in
redshift space:
\begin{equation}
\label{eqn:wp}
\wrp = 2 \int_0^{\pi_{\rm max}} \xi(r_p,\pi)  d\pi,
\end{equation}
where $r_p$ and $\pi$ are separations transverse and parallel to the line of
sight.  We adopt $\pi_{\rm max}=40 h^{-1}{\rm Mpc}$ (in the measurement and
modeling), large enough to include nearly all correlated pairs and thus
minimize redshift-space distortion while keeping background noise from
uncorrelated pairs low.  Because our sample is volume-limited, we are measuring
the clustering of a homogeneous population of galaxies throughout the survey
volume, which greatly simplifies HOD modeling.  Further details of the sample and
measurement are given in Z04b.  In our analysis, we use 11 $\wrp$ data points
in the range $0.1 h^{-1}{\rm Mpc}< r_p < 20 h^{-1}{\rm Mpc}$, sufficiently
below the projection scale $\pi_{\rm max}$ to avoid contamination of redshift
space distortions, though we have found that including points up to $r\approx
40 h^{-1} \rm Mpc$, which have low statistical weight, does not alter our
results.

We also require our models to reproduce the measured mean comoving space
density of our sample, $\bar{n}_g^{\rm obs}=1.17 \times 10^{-3} h^3 {\rm
  Mpc}^{-3}$.  This quantity has an uncertainty due to sample variance that can
be written as $\sigma_{\bar{n}}/\bar{n}_g^{\rm obs} =
\sqrt{\left<\delta^2_g\right>}$, where $\left<\delta^2_g\right>$ is the
variance of the galaxy overdensity.  We estimate $\left<\delta^2_g\right>$ by
integrating the two-point correlation function over the volume of the SDSS
$\Mr$ sample.  To compute this integral we generate a large number of
independent random pairs of points within the sample volume and sum $\xi(r)$
over all these pairs.  We use the Z04b correlation function and extend it to
larger scales with the linear theory correlation function multiplied by $b^2$,
where $b=1.4$ is the large-scale bias factor for $\Mr$ galaxies.  We vary the
number of random pairs used and find that the integral converges at $10^5$
pairs.  Using two million pairs, we find that the number density uncertainty
due to sample variance is $\sigma_{\bar{n}}/\bar{n}_g^{\rm obs} =
0.0377$. There is also a shot noise Poisson uncertainty in the number density
that, for this number of galaxies, is $0.0062$.  We add these two components in
quadrature to obtain a total uncertainty of $\sigma_{\bar{n}}^{\rm
  tot}/\bar{n}_g^{\rm obs} = 0.0382$, and therefore
\begin{equation}
\bar{n}_g^{\rm obs}=(1.17 \pm 0.05) \times 10^{-3} h^3 {\rm Mpc}^{-3}.
\end{equation}

\section{Results}
\label{section:results}

\begin{deluxetable*}{lllllll}
\tablecaption{Cosmological plus HOD parameters, marginalized constraints with
  $68.3\%$ C.L. errors. \label{table}}
\tablewidth{0pt}
\tablehead{
\colhead{Parameter}
& \colhead{CMB+$w_p(r_p)$ $3p$+$P_g(k)$} 
& \colhead{CMB+$w_p(r_p)$ $3p$} 
& \colhead{CMB+$w_p(r_p)$ $4p$} 
& \colhead{CMB+$w_p(r_p)$ $5p$} &
\colhead{CMB+$P_g(k)$}  & \colhead{CMB} 
}
\startdata
$A$            
& $0.731^{+0.057}_{-0.053}$    
& $0.749^{+0.063}_{-0.058}$    
& $0.768^{+0.074}_{-0.068}$  
& $0.803^{+0.103}_{-0.093}$ 
& $0.747^{+0.077}_{-0.071}$    
& $0.79^{+0.15}_{-0.12}$
\\[0.025 in] 
$n$            
& $0.947^{+0.017}_{-0.018}$  
& $0.953^{+0.019}_{-0.019}$  
& $0.958^{+0.020}_{-0.021}$  
& $0.968^{+0.027}_{-0.027}$
& $0.956^{+0.020}_{-0.021}$ 
& $0.974^{+0.037}_{-0.036}$ 
\\[0.025 in] 
$\tau$         
& $0.100^{+0.017}_{-0.021}$ 
& $0.115^{+0.019}_{-0.023}$ 
& $0.131^{+0.021}_{-0.027}$  
& $0.155^{+0.026}_{-0.037}$
& $0.105^{+0.017}_{-0.028}$ 
& $0.158^{+0.093}_{-0.084}$
\\[0.025 in] 
$h$            
& $0.687^{+0.023}_{-0.023}$  
& $0.698^{+0.026}_{-0.026}$  
& $0.708^{+0.028}_{-0.029}$  
& $0.721^{+0.034}_{-0.036}$ 
& $0.697^{+0.028}_{-0.028}$ 
& $0.755^{+0.059}_{-0.058}$  
\\[0.025 in] 
$\Omega_c h^2$ 
& $0.1148^{+0.0039}_{-0.0039}$  
& $0.1126^{+0.0043}_{-0.0044}$  
& $0.1107^{+0.0050}_{-0.0049}$  
& $0.1088^{+0.0058}_{-0.0057}$
& $0.1176^{+0.0069}_{-0.0069} $  
& $0.104^{+0.013}_{-0.013}$  
\\[0.025 in] 
$\Omega_b h^2$ 
& $0.02234^{+0.00079}_{-0.00080}$  
& $0.02247^{+0.00084}_{-0.00084}$  
& $0.02263^{+0.00086}_{-0.00087}$  
& $0.0229^{+0.0010}_{-0.0010}$
& $0.0227^{+0.0009}_{-0.0009} $  
& $0.0230^{+0.0013}_{-0.0013}$  
\\[0.025 in] 
$M_1 [10^{13} h^{-1}M_\odot]$          
& $4.79^{+0.63}_{-0.63}$
& $4.58^{+0.60}_{-0.60}$
& $4.52^{+0.63}_{-0.63}$
& $3.31^{+1.61}_{-1.64}$
&$-$ 
&$-$  
\\[0.025 in] 
$M_{\rm min}[10^{12} h^{-1} M_\odot]$  
& $3.23^{+0.36}_{-0.35}$
& $3.03^{+0.36}_{-0.36}$
& $3.26^{+0.46}_{-0.48}$
& $3.08^{+0.49}_{-0.51}$
&$-$ 
&$-$ 
\\[0.025 in] 
$\kappa$         
& $4.44^{+0.48}_{-0.63}$ 
& $4.44^{+0.51}_{-0.69}$ 
& $3.85^{+0.45}_{-0.56}$  
& $6.31^{+1.33}_{-1.94}$ 
&$-$ 
&$-$  
\\[0.025 in] 
$\sigma_{\rm cen}$          
&$0$ 
&$0$ 
& $0.41^{+0.11}_{-0.18}$  
& $0.39^{+0.10}_{-0.18}$ 
&$-$ 
&$-$  
\\[0.025 in] 
$\alpha$       
& $1$ 
& $1$ 
& $1$ 
& $0.83^{+0.22}_{-0.23}$
& $-$ 
& $-$
\\[0.025 in] 
$b_g$$^*$
& $1.48^{+0.088}_{-0.088}$
& $1.47^{+0.093}_{-0.093}$
& $1.43^{+0.077}_{-0.077}$
& $1.40^{+0.095}_{-0.095}$  
& $-$
& $-$
\\[0.025 in] 
$\Omega_m$     
& $0.292^{+0.025}_{-0.024}$
& $0.278^{+0.027}_{-0.027}$
& $0.268^{+0.029}_{-0.029}$
& $0.256^{+0.034}_{-0.033}$  
& $0.291^{+0.034}_{-0.034}$ 
& $0.231^{+0.054}_{-0.056}$
\\[0.025 in]
$\sigma_8$     
& $0.809^{+0.028}_{-0.028}$ 
& $0.812^{+0.028}_{-0.027}$ 
& $0.816^{+0.030}_{-0.030}$ 
& $0.829^{+0.039}_{-0.039}$  
& $0.834^{+0.049}_{-0.050}$
& $0.802^{+0.072}_{-0.073}$
\\[0.025 in] 
$\chi^2_{\rm eff}/\rm DOF$ 
& $1483.4/1391$
& $1458.5/1373$
& $1458.4/1372$ 
& $1458.1/1371$ 
& $1477.2/1383$ 
& $1452.5/1365$
\\[0.025 in] 
AIC
& $1503.4$
& $1476.5$
& $1478.4$
& $1480.1$
& $1491.2$
& $1464.5$
\\
BIC
& $1555.9$
& $1523.5$
& $1530.7$ 
& $1537.7$ 
& $1527.9$ 
& $1495.8$
\enddata
\tablecomments{$^*$The large scale galaxy bias ($k\rightarrow 0$).}
\end{deluxetable*}

\begin{figure}
\plotone{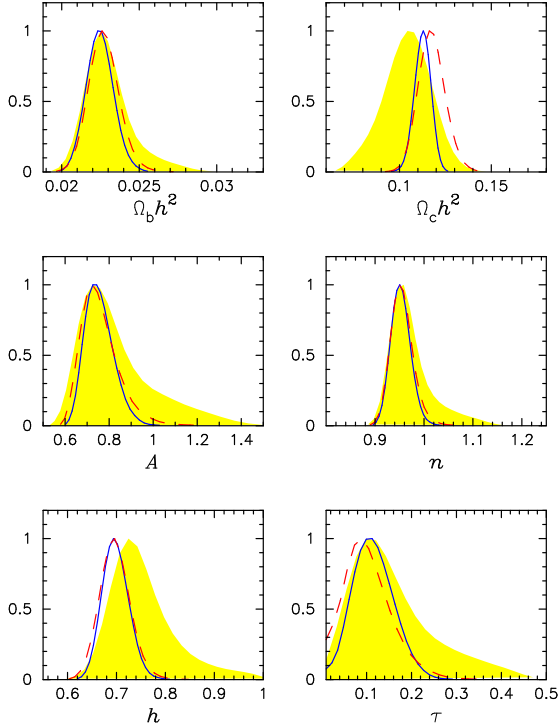}
\caption{Shown are the marginalized posterior likelihoods for the cosmological
  parameters determined from CMB+$w_p(r_p)[M_{^{0.1}r}<-21]$ with a
  three-parameter HOD in solid blue, that for CMB+SDSS 3D $P_g(k)$ in red dashed,
  and that for the CMB alone in yellow (gray) shaded. \label{parameters1}}
\end{figure}

For a given cosmology and HOD parameter choice, we use the predicted $w_p(r_p)$
to calculate the likelihood to observe the $M_{^{0.1}r} < -21$ sample's
$w_p(r_p)$ and $\bar{n}_g^{\rm obs}$. We combine this likelihood with that for
the model's prediction for the cosmic microwave background anisotropy
temperature correlation and temperature-polarization cross-correlation to
produce the WMAP (first year), ACBAR ($\ell >800$), CBI ($600<\ell <2000$) and VSA
($\ell >600$) observations
(\citealt{hinshaw03,verde03,kogut03,dickinson04,readhead04}).  We vary the six
parameters for the ``vanilla'' $\Lambda$CDM cosmological model plus the five
HOD parameters: $\mathbf p = (\Omega_b h^2, \Omega_c
h^2,\Theta_s,\ln(A),n,\tau,\Mmin,M_1,\kappa,\sigma_{\rm cen},\alpha)$.  The ranges
allowed in our Markov Chain Monte Carlo (MCMC) sampling of parameters are
chosen to avoid any artificial cut-off of the likelihood space and are
\begin{eqnarray}
0.005\le& \Omega_b h^2 &\le 0.1\cr
0.01\le& \Omega_c h^2 &\le 0.99\cr
0.005 \le&  \Theta_s &\le 0.1\cr
-0.68 \le& \ln(A) &\le 0.62\cr
0.5 \le& n &\le 1.5\cr
0.01 \le& \tau &\le 0.8\cr
10^9\, M_\odot \le& \Mmin &\le 5\times 10^{13}\, M_\odot \cr
10^{10}\, M_\odot \le& M_1 &\le 4\times 10^{14}\, M_\odot \cr
1 \le& \kappa &\le 30\cr
0  \le& \sigcen &\le 10\cr
0.2  \le& \alpha &\le 2.5\; .
\end{eqnarray}
Here, $A$ is related to the amplitude of curvature fluctuations at horizon
crossing, $|\Delta R|^2 = 2.95\times 10^{-9} A$ at the scale $k= 0.05 \rm \, Mpc$.  The
angular acoustic peak scale $\Theta_s$ is the ratio of the sound horizon at last
scattering to that of the angular diameter distance to the surface of last
scattering \citep{Kosowsky:2002zt}. 

To measure the likelihood space allowed by the data, we use a Metropolis MCMC
method with a modified version of the \citet{lewis02} CosmoMC code.  We use the
WMAP team's code to calculate the WMAP first-year observations' likelihood, and
CosmoMC to calculate that for ACBAR, CBI and VSA. After burn-in, the chains
typically sample $10^5$ points, and convergence and likelihood statistics are
calculated from these.  Since it is not known {\it a priori} which HOD
parameters are most constrained by the $\wrp$ measurement, we use the Akaike
and Bayesian Information Criteria (AIC and BIC) to determine which parameters
are statistically relevant to describing $\wrp$ (\citealt{Akaike74,Schwarz78};
see also \citealt{liddle04}).  More parameters might well be needed once we
have more data to constrain the HOD, but $\wrp$ alone doesn't provide enough
information to demand it.

Likelihood analyses were performed for several cas\-es where some parameters
were kept free and others were fixed to a physical limit, i.e. where the
scatter in the mass-luminosity relation is unimportant ($\sigma_{\rm cen}\equiv
0$), the cut-off scale of the satellite galaxies is exactly that of the minimum
mass $\Mmin$ ($\kappa\equiv 1$), or to a value ($\alpha\equiv 1$) predicted in
the numerical simulations and semi-analytic models of satellite halo
distributions in \citet{berlind03} and \citet{kravtsov04}.  
If we adopt all three of these constraints and allow only two parameters,
$\Mmin$ and $M_1$, to vary to fit $\wrp$ and $\bar{n}_g^{\rm obs}$, then we
obtain a poor fit.  This model is an inadequate description of the data
according to the information criteria ($\Delta{\rm BIC}=7.2$ and $\Delta{\rm
  AIC}=12.5$) relative to the three-parameter $\Mmin$, $M_1$, and $\kappa$
model ($3p$).  We also investigated a four parameter model ($4p$), varying
$\Mmin$, $M_1$, $\kappa$, and $\sigcen$ with $\alpha\equiv 1$, as well as a
five parameter model ($5p$) varying all parameters in this HOD.  Relative to
the $3p$ model, the $4p$ and $5p$ models introduce new parameters that are not
justified by the information criteria ($\Delta {\rm BIC}>6$,
cf. Table~\ref{table}), since these models add freedom but yield only a very
small reduction in $\chi^2$.

To assess the importance of one aspect of the halo modeling, we performed a
test on the $3p$ model whereby the NFW concentration $c_0$ [Eq.~(\ref{c0})] of
dark matter is replaced by that for the galaxies, $c_0^{\rm gal}$, and is also
left free in the MCMC within $0.01 < c_0^{\rm gal} < 200$ and independent of
the dark matter concentration of the halos.  We find that the derived
cosmological parameters and their uncertainties remain nearly unchanged from a
model with no spatial bias, and the constraints on the galaxy concentration are
consistent with no spatial bias: $c_0^{\rm gal} = 11.1^{+0.7}_{-5.3}$.  The
marginalized values of the HOD parameters $\Mmin$ and $M_1$ remain unchanged
with varying $c_0^{\rm gal}$, though the error on the cut-off scale of the
satellite galaxies $\kappa$ increases ($\kappa =4.71^{+0.60}_{-1.41}$).  This
increase is expected since it is precisely the central distribution of
satellite galaxies that is positively correlated to the one-halo galaxy
distribution concentration, with a correlation coefficient of $r=0.72$.  As a
large $c_0^{\rm gal}$ makes the distribution of galaxies inside halos more
concentrated, to maintain the small-scale clustering, $\kappa$ increases to
allow relatively more galaxies to be put in halos with larger virial radii and
lower concentrations.

\begin{figure}
\plotone{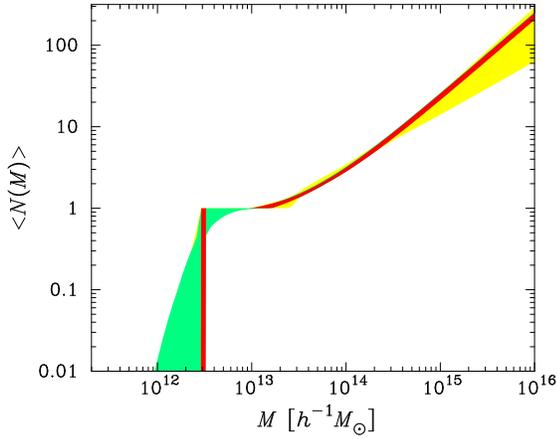}
\caption{Plotted are the $\Delta\chi^2 < 1$ range from the best fit for the HOD
  drawn from the MCMC chains for the $3p$ model in red (dark gray), $4p$ model
  in green (medium gray), and $5p$ model in yellow (light gray).
  \label{parameters2}}
\end{figure}

\begin{figure}
\plotone{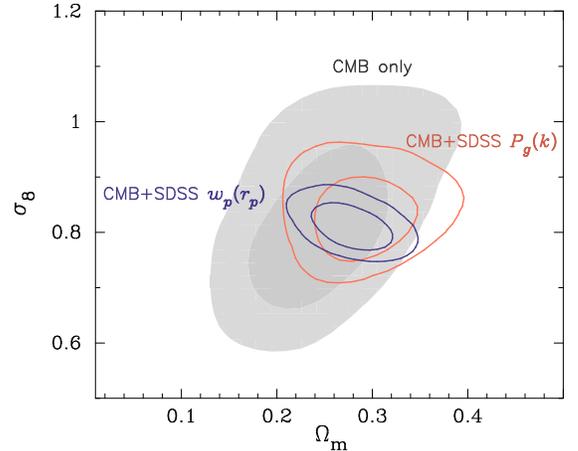}
\caption{Shown are the marginalized 68.3\% and 95.4\% C.L. contours in $\sigma_8$
 vs. $\Omega_m$ for the WMAP+ACBAR+CBI+VSA (CMB) data alone (gray shaded), from
 the CMB + SDSS 3D $P_g(k)$ (orange/light-gray lines) and CMB + SDSS $w_p(r_p)$
 (blue/dark-gray lines) from the $3p$ HOD analysis presented here.
 \label{parameters2D}}
\end{figure}

Figure 1 illustrates the way that $\wrp$ constrains cosmological parameters.
Data points show the Z04b measurements, and the solid line shows the prediction
of the best-fit $3p$ model.  Dashed curves show the prediction of the $\wrp$
after $\sigma_8$ is perturbed by $\pm 3\sigma$ relative to its best-fit value
given in column 2 of Table~1, with all other cosmological parameters (and
therefore the shape of the linear matter power spectrum) as well as the HOD
parameters held fixed.  Dotted curves show the prediction of $\wrp$ after
changing $\Omega_c$, and thus the shape of the transfer function in the matter
power spectrum, by $\pm 3\sigma$, with all other cosmological and HOD
parameters fixed.  The strength of the constraints derived from $\wrp$ stems
from the combined relative dependence of the 1-halo and 2-halo regimes and
therefore the overall shape of $\wrp$.

The cosmological parameters' marginalized posterior likelihoods for the $3p$
model are shown in Figure~\ref{parameters1}. Also shown for comparison are the
marginalized likelihoods for the CMB plus SDSS 3D $P_g(k)$ [updated from
  \citet{tegmark04} with new CMB results], and that from the CMB data alone.
We also combine the CMB+$\wrp (3p)$ measurement with the SDSS 3D $P_g(k)$ for a
joint constraint on cosmological parameters.  Since the $P_g(k)$ data points
included in the analysis are at wavelengths $\lambda=2\pi/k > 30\, h^{-1}\,{\rm
  Mpc}$, the information they contain is largely independent of that in the
$\wrp$ data points at $r_p < 20\, h^{-1}\,{\rm Mpc}$.  All parameters' best fit
values and errors are listed in Table~\ref{table}.  The resulting range of the
HOD measured for all models here are shown in Figure~\ref{parameters2}.

Two-dimensional contours of $\Omega_m$ and $\sigma_8$ are shown in
Figure~\ref{parameters2D} and are compared to those obtained using CMB alone or
CMB + $P_g(k)$. The anticorrelation of $\Omega_m$ and $\sigma_8$ from the
$\wrp$ constraint seen in Fig.~\ref{parameters2D} arises from the
anticorrelated degeneracy in these parameters in the one-halo component due to
its dependence on the halo mass function which needs to maintain its amplitude
at high halo masses, and the $Omega_m$ and $\sigma_8$ anticorrelation in the
two-halo component due to the amplitude-shape degeneracy of the dark matter
power spectrum (or dark matter correlation function).

Important results to note from the Figures and Table are the following.
Cosmological constraints obtained using CMB and $\wrp$ are substantially
tighter than those from CMB alone, and they are similar in value and tightness
to those obtained from CMB + $P_g(k)$ despite the introduction of new
parameters to represent the HOD.  The $\sigma_8$ constraints using $\wrp$ are
tighter than those using $P_g(k)$; note that the latter estimate has dropped
relative to that of \cite{tegmark04} because of the smaller scale CMB data.  If
we incorporate $P_g(k)$ constraints in addition to $\wrp$, then parameter
values change by less than $1-\sigma$ and error bars improve slightly.  Our
cosmological parameter results also agree, within errors, with the recent
results from SDSS galaxy bias and Lyman-$\alpha$ forest \citep{seljakparams04}.
The HOD parameters are partially degenerate among themselves, so adding
parameters to the HOD model worsens the constraint on any one of them.
However, within the range of models examined here, adding parameters to the HOD
only slightly increases the error bars on cosmological parameters.

\begin{figure}
\plotone{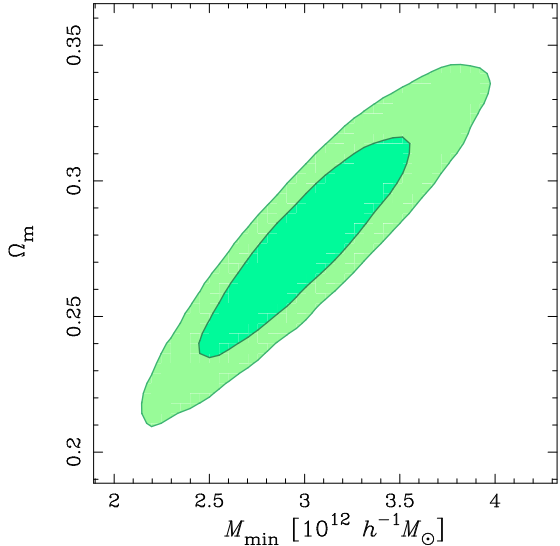}
\caption{Plotted are the 68.3\% and 95.4\% C.L.\ contours for the marginalized
  likelihoods for $\Omega_m$ vs. $M_{\rm min}$.  The strong degeneracy
  (correlation of $r=0.92$) roughly follows $\Omega_m\propto M_{\rm min}$, as
  expected \citep{zheng02,rozo04}
  \label{om_degen}}
\end{figure}

Since the small scales of the primordial power spectrum probed by $w_p(r_p)$
could be useful in constraining any deviations from a simple power-law
primordial spectrum as well as a model including the suppression in power
spectrum and mass function due to the presence of massive neutrinos, we
performed an MCMC analysis including a running of the spectrum $dn/d\ln k$
about the scale $k=0.05\rm\;Mpc$ for the $3p$ HOD model as well a model
including massive neutrinos.  We find little evidence for running, $dn/d\ln k =
-0.062^{+0.026}_{-0.027}$, comparable to the results of \citet{spergel03}.  The
halo model in the presence of massive neutrinos is applied such that the halo
mass function, halo bias and halo profile is that of the cold dark matter alone
since neutrino clustering is a very small effect on these quantities
\citep{Abazajian:2004zh}. The presence of massive neutrinos is constrained to
$m_\nu < 0.27\rm\,eV$ ($95\%$ C.L.) for each of 3 neutrinos with degenerate
mass.  The statistical errors from our analysis of the CMB plus this $\wrp$
measurement on $dn/d\ln k$ and $m_\nu$ are comparable to those from other
cosmological parameter analyses, being smaller than those from the shape of the
SDSS 3D $P_g(k)$ plus WMAP \citep{tegmark04}, comparable to the WMAP plus
2dFGRS 3D $P_g(k)$ plus modeled bias constraints \citep{spergel03}, but not as
stringent as those from modeling the galaxy bias in the SDSS from galaxy-galaxy
lensing and clustering of the Lyman-$\alpha$ forest in the SDSS
\citep{seljaklens04,seljakparams04}.

\section{Discussion}

The remaining uncertainties in cosmological parameters introduce relatively
little uncertainty in the HOD parameters, i.e., we now know the underlying
cosmology with sufficient precision to pin down the relation between galaxies
and mass.  The strongest expected degeneracy is between the value of $\Omega_m$
and the mass scale parameters $M_{\rm min}$ and $M_1$, since one can compensate
a uniform increase in halo masses by simply shifting galaxies into more massive
halos \citep{zheng02,rozo04}.  The error contours for $\Omega_m$ vs.\ $M_{\rm
  min}$ are shown in Figure~\ref{om_degen}. The degeneracy between these
parameters is strong, with a correlation of $r=0.96$.  While this degeneracy
would cause large uncertainties in the values of $M_{\rm min}$ and $\Omega_m$
if we used the galaxy clustering data alone, the combination of CMB and $\wrp$
data constrains $\Omega_m$ fairly tightly, leaving limited room to vary the
mass scale parameters.  Incorporating SDSS clustering measures that are
directly sensitive to halo masses, such as redshift-space distortions
\citep{zehavi02} and galaxy-galaxy lensing measurements
\citep{sheldon04,seljaklens04}, may further improve the $\Omega_m$ constraints.

As discussed by \cite{berlind02}, the galaxy correlation function places
important constraints on HOD parameters, but it still allows tradeoffs between
different features of $P(N|M)$ and (to a lesser degree) between $P(N|M)$ and
the assumed spatial bias of galaxies within halos.  Additional clustering
statistics such as the group multiplicity function, higher order correlation
functions, and void probabilities impose complementary constraints that can
break these degeneracies.  Our analysis should thus be seen as a first step in
a broader program of combining galaxy clustering measurements from the SDSS and
other surveys with other cosmological observables to derive simultaneous
constraints on cosmological parameters and the galaxy HOD [see
  \citet{berlind02,weinberg02,zheng02} for further discussion].
\citet{vandenbosch03} have been carrying out a similar program using the
closely related conditional luminosity function (CLF) method applied to the
2dFGRS luminosity and correlation functions (see also
\citealt{vandenbosch03err}).  They find $\sigma_8 = 0.78\pm 0.12$ and $\Omega_m
= 0.25^{+0.10}_{-0.07}$ in their analysis combined with CMB data, with both
errors at $95\%$ C.L. as given in that work.  Our results are in agreement,
within errors, with their determinations of $\sigma_8$ and $\Omega_m$.

Besides the statistical error bars, there are two main sources of systematic
uncertainty in our cosmological parameter estimates.  The first is the
possibility that our HOD parameterization does not have enough freedom to
describe the real galaxy HOD, and that we are artificially shrinking the
cosmological error bars by adopting a restrictive theoretical prior in our
galaxy bias model.  For the $3p$ model, this is arguably the case, since it
effectively assumes perfect correlation between the mass of a halo and the
luminosity of its central galaxy.  However, our $5p$ model is able to give an
essentially perfect description of the predictions of semi-analytic galaxy
formation models and hydrodynamic simulations (\citealt{zheng04}; see also
\citealt{guzik02,berlind03,kravtsov04}), so there is good reason to think that
the error bars quoted for this case are conservative.  This model still makes
the assumption that satellite galaxies have no spatial bias with respect to
dark matter within halos, but the concentration test in \S\ref{section:results}
shows that dropping this assumption has minimal impact on cosmological
conclusions.  In place of an HOD model, traditional analyses based on the
large-scale galaxy power spectrum assume that the galaxy power spectrum is a
scale-independent multiple of the linear matter power spectrum, so that their
shapes are identical.  Scale-independence in the linear regime is expected on
fairly general grounds
\citep{coles93,fry93,weinberg95,mann98,scherrer98,narayanan00}.  However, it is
not clear just how well this approximation holds over the full range of scales
used in the power spectrum analyses, so although our HOD models are
considerably more complex than linear bias models, our approach is arguably no
more dependent on theoretical priors.  In future work, we can use the HOD
modeling to calculate any expected scale-dependence of the power spectrum bias,
thus improving the accuracy of the power spectrum analyses and allowing them to
extend to smaller scales.

The second source of systematic uncertainty is the possibility that our
approximation for calculating $\wrp$ for a given cosmology and HOD is
inaccurate in some regions of our parameter space.  The ingredients of this
approximation have been calibrated or tested on N-body simulations of
cosmological models similar to the best fitting models found here, so we do not
expect large inaccuracies.  However, there are several elements of the halo
model calculation that could be inaccurate or cosmology dependent at the 10\%
level that is now of interest, including departures from the \cite{jenkins01}
mass function, scale dependence of halo bias, and details of halo exclusion.
Uncertainties in the halo mass-concentration relation and the impact of scatter
in halo concentrations come in at a similar level, though the test in
\S\ref{section:results} again indicates that these uncertainties mainly affect
the details of the derived HOD, not the cosmological parameter determinations.
Without a comprehensive numerical study of these issues, it is difficult to
assess how large the systematic effects on our parameter determinations could
be, but we would not be surprised to find that they are comparable to our
statistical errors.  We plan to carry out such a study to remove this source of
systematic uncertainty in future work; the papers of \cite{seljak04} and
\cite{Tinker:2004gf} present steps along this path.

Analyses of multiple classes of galaxies will allow consistency checks on any
cosmological conclusions, since different classes will have different HODs but
should yield consistent cosmological constraints.  By drawing on the full range
of galaxy clustering measurements, joint studies of galaxy bias and
cosmological parameters will sharpen our tests of the leading theories of
galaxy formation and the leading cosmological model.  With this current
analysis alone, we find that the combination of CMB anisotropies and
small-scale galaxy clustering measurements provides, simultaneously, tight
constraints on the occupation statstics of galaxies in dark matter halos, and
some of the best available constraints on fundamental cosmological parameters.

\acknowledgments 

We thank Salman Habib, Katrin Heitmann, Wayne Hu, Andrey Kravtsov, Chung-Pei
Ma, Peder Norberg, Roman Scoccimarro, Uro\v{s} Seljak, Erin Sheldon, Lu\'{\a
  i}s Teodoro, Jeremy Tinker, Roberto Trotta, Frank van den Bosch, Risa
Wechsler and Martin White for fruitful discussions.

Funding for the creation and distribution of the SDSS Archive has been provided
by the Alfred P. Sloan Foundation, the Participating Institutions, the National
Aeronautics and Space Administration, the National Science Foundation, the
U.S. Department of Energy, the Japanese Monbukagakusho, and the Max Planck
Society. The SDSS Web site is \url{http://www.sdss.org/}.

The SDSS is managed by the Astrophysical Research Consortium (ARC) for the
Participating Institutions. The Participating Institutions are The University
of Chicago, Fermilab, the Institute for Advanced Study, the Japan Participation
Group, The Johns Hopkins University, the Korean Scientist Group, Los Alamos
National Laboratory, the Max-Planck-Institute for Astronomy (MPIA), the
Max-Planck-Institute for Astrophysics (MPA), New Mexico State University,
University of Pittsburgh, Princeton University, the United States Naval
Observatory, and the University of Washington.

\bibliography{hod}
\bibliographystyle{apj}

\end{document}